\documentstyle[aps,epsfig,float,amstex]{revtex}

\title{Time-dependent quantum transport: an exact formulation based on TDDFT}

\author{Gianluca Stefanucci and Carl-Olof Almbladh}

\address{
Department of Solid State Theory, 
Institute of Physics, 
Lund University,\\
S\"olvegatan 14 A, 223 62 Lund, Sweden}

\begin{document}
    
\def\tg{\mbox{\textsl{g}}}
\def\bS{\mbox{\boldmath $\Sigma$}}
\def\bDelta{\mbox{\boldmath $\Delta$}}
\def\bcalE{\mbox{\boldmath ${\cal E}$}}
\def\bcalG{\mbox{\boldmath ${\cal G}$}}
\def\bcalH{\mbox{\boldmath ${\cal H}$}}
\def\bcalV{\mbox{\boldmath ${\cal V}$}}
\def\bcalO{\mbox{\boldmath ${\cal O}$}}
\def\bcalQ{\mbox{\boldmath ${\cal Q}$}}
\def\bcalS{\mbox{\boldmath ${\cal S}$}}
\def\bG{\mbox{\boldmath $G$}}
\def\bg{\mbox{\boldmath $g$}}
\def\gC{\mbox{\boldmath $C$}}
\def\gZ{\mbox{\boldmath $Z$}}
\def\gR{\mbox{\boldmath $R$}}
\def\gN{\mbox{\boldmath $N$}}
\def\bV{\mbox{\boldmath $V$}}
\def\ua{\uparrow}
\def\da{\downarrow}
\def\a{\alpha}
\def\b{\beta}
\def\g{\gamma}
\def\G{\Gamma}
\def\d{\delta}
\def\D{\Delta}
\def\e{\epsilon}
\def\ve{\varepsilon}
\def\z{\zeta}
\def\h{\eta}
\def\th{\theta}
\def\k{\kappa}
\def\l{\lambda}
\def\L{\Lambda}
\def\m{\mu}
\def\n{\nu}
\def\x{\xi}
\def\X{\Xi}
\def\p{\pi}
\def\P{\Pi}
\def\r{\rho}
\def\s{\sigma}
\def\S{\Sigma}
\def\t{\tau}
\def\f{\phi}
\def\vf{\varphi}
\def\F{\Phi}
\def\c{\chi}
\def\w{\omega}
\def\W{\Omega}
\def\Q{\Psi}
\def\q{\psi}
\def\de{\partial}
\def\inf{\infty}
\def\ra{\rightarrow}
\def\bra{\langle}
\def\ket{\rangle}

\maketitle

\begin{abstract}
An exact theoretical framework based on Time Dependent Density Functional Theory 
(TDDFT) is proposed in order to deal with the time-dependent quantum transport in 
fully interacting systems. We use a 
\textit{partition-free} approach by Cini in which the 
whole system is in equilibrium 
before an external electric field is switched on. 
Our theory includes the interactions 
between the leads and between the leads and the device.
It is well suited for calculating measurable transient phenomena as well as
a.c. and other time-dependent responses.
We show that the steady-state current results from a \textit{dephasing 
mechanism} provided the leads are macroscopic and the device is 
finite. In the d.c. case, we obtain a Landauer-like formula when 
the effective potential of TDDFT is uniform deep inside the electrodes.
\end{abstract}

\section{Introduction}
\begin{figure}[H]
\begin{center}
	\epsfig{figure=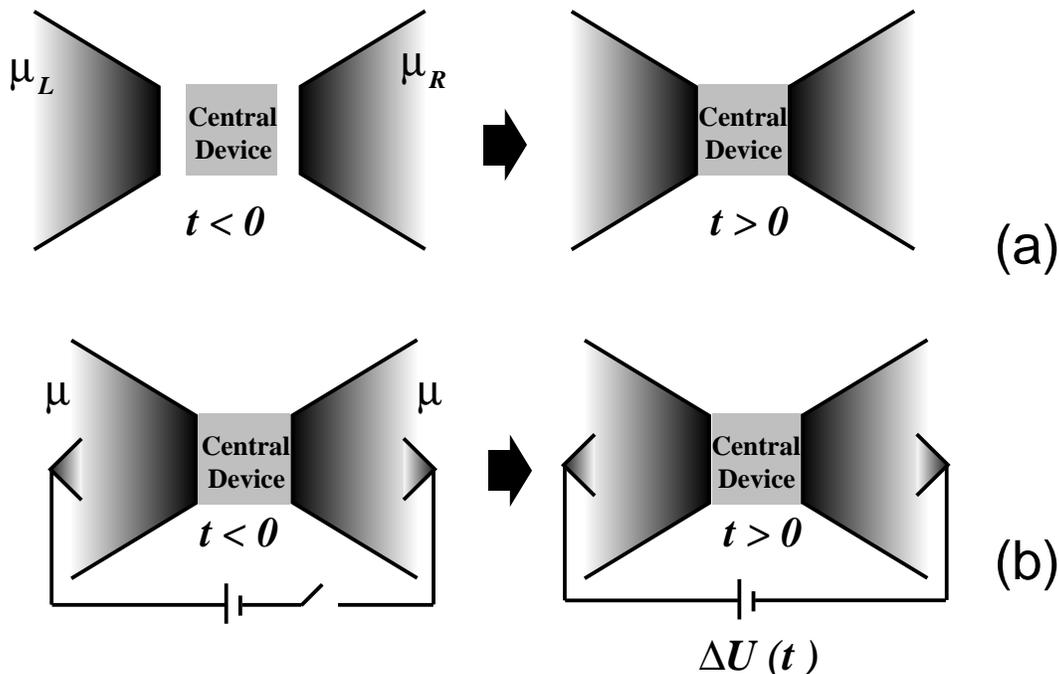,width=14cm}
	\caption{\footnotesize{(a) Schematic representation of the 
partitioned approach by Caroli \textit{et al.} On the left the electrodes 
are disconnected and in thermodynamic equilibrium at two different chemical 
potential. On the right the contacts are established and electrons start to flow. 
(b) Schematic representation of the partition-free approach by Cini. The 
whole system is in thermodynamic equilibrium and it is characterised 
by a unique chemical potential. Then, a bias is applied deep inside 
the electrodes.}}
\label{sm}
\end{center}
\end{figure}

During the last decade, the experimental progress in 
the fabri\-ca\-tion/ma\-ni\-pulation of 
nano-systems like quantum wires and dots has enhanced the interest in 
the theory of quantum transport. 
The Landauer-B\"uttiker formalism\cite{landauer,buttiker} applies to 
noninteracting systems and gives the steady-state current of 
macroscopic electrodes connected via mesoscopic (or nanoscopic) 
constrictions. Techniques based on non-equilibrium Green 
functions\cite{kadabaym,keldysh} provide a natural 
framework for obtaining the Landauer formula starting from a microscopic theory. 
The first attempt was made by Caroli \textit{et al.} \cite{caroli1,caroli2} 
who consider the two 
leads as isolated subsystems with different chemical potentials 
in the remote past. The current will flow through 
the system once the contacts between the device and the leads have  
been established, see Fig. \ref{sm}(a). This approach is based on a fictitious 
partitioning since in a real experiment the whole system is in thermodynamic 
equilibrium before an external bias is applied deep inside the 
electrodes. Later on 
this partitioned scheme was adopted by Meir and Wingreen\cite{meir} 
to obtain the steady-state current through an \textit{interacting}
central device. Shortly after, Wingreen \textit{et al.}
\cite{wingreen,jauho,haug} generalised the  Meir-Wingreen formula  
to time-dependent phenomena. 

Despite the important results mentioned above, the partitioned scheme 
has several drawbacks. First, it is difficult to include the interactions 
between the leads and between the leads and the device. These 
interactions are responsible for the establishing of dipole layers and charge 
transfers which shape the potential landscape in the device region 
and influence the contacts. 
Second, there is a crucial assumption of equivalence 
between the long-time behaviour of the 1) initially partitioned and 
biased system once the \textit{coupling} between the subsystems is  
switched on and 2) the whole partition-free system  
once the \textit{bias} is applied. Third, 
there is no well defined prescription which fixes the initial 
equilibrium distribution of the isolated central device; this makes 
the transient current difficult to interpret.

A conceptually different approach from the one by Caroli \textit{et al.} 
has been introduced by Cini\cite{cini}, see Fig. \ref{sm}(b). Here, 
the system is contacted and in thermodynamic equilibrium before an 
external time-dependent disturbance is switched on. Cini developed 
the general theory for the case of free electrons with his focus on 
semiconductor junction devices. 
In this letter we  extend the Cini theory to 
fully \textit{interacting} (electrodes and device) systems. 
Our exact theoretical framework is based on Time Dependent Density 
Functional Theory (TDDFT) and is well suited to describe the 
steady state as well as a.c. and other \textit{time-dependent} 
current responses.

\section{General formulation}
To be specific let us focus on the Coulomb interaction and on paramagnetic 
systems. In the case we only ask for the time-dependent  
density $n({\bf r},t)$ the original Density Functional 
Theory\cite{hk,ks} and its finite-temperature generalisation\cite{mermin}  
has been extended to time-dependent phenomena \cite{rg,litong}. 
This theory applies only to those cases where the external disturbance 
is local in space, $\int d{\bf r}\;U({\bf r},t) n({\bf r},t)$. Runge 
and Gross\cite{rg} have shown that one can compute $n({\bf r},t)$ in a one-particle 
manner using an effective potential
$$
U^{\rm eff}({\bf r},t)=U_{\rm C}({\bf r},t)+
v_{\rm xc}({\bf r},t).
$$
Here, $U_{\rm C}=U+V_{\rm H}+V_{\rm n}$ is the classical 
electrostatic potential and it is given by the sum of the external potential 
$U$, the Hartree potential $V_{\rm H}$ and the Coulomb potential from 
the nuclei $V_{\rm n}$. Furthermore, $v_{\rm xc}$ accounts for exchange and 
correlations and is obtained from an exchange-correlation action functional, 
$v_{\rm xc}({\bf r},t)=\d A_{\rm xc}[n]/\d n({\bf r},t)$  
(as pointed out in Ref. \cite{vanleeuwen}, the causality 
and symmetry properties require that the action functional 
$A_{\rm xc}[n]$ is defined on the Keldysh contour). Thus, the 
effective one-body Hamiltonian reads $
\bcalH^{\rm eff}(t)=\int d {\bf r}\;\left[
\left(-\frac{\nabla^{2}}{2m}\right)+
U^{\rm eff}({\bf r},t)
\right]$. Here and in what follows, we use boldface for operators in  
one-particle Hilbert space, and $\hbar$ is 1 in our units.

It is convenient to define the projectors 
$P_{\a}=\int_{\a}d {\bf r}|{\bf r}\ket\bra {\bf r}|$ onto the left or right 
leads ($\a=L,R$) or the central device ($\a=D$) and to introduce the notation 
$\bcalO_{\a\b}\equiv P_{\a}\bcalO P_{\b}$, where $\bcalO$ is an arbitrary operator 
in one-body space \cite{foot}. 
The one-particle scheme of Time Dependent Density Functional Theory 
(TDDFT) corresponds to a fictitious Green function 
$\bcalG(t;t')$ which satisfies a one-particle equation of motion 
\begin{equation}
\left\{i\frac{d}{dt}-\bcalE(t)-\bV\right\}\bcalG(t;t')=\d(t-t').
\label{eomg}
\end{equation}
In eq. (\ref{eomg}) $\bcalE=\bcalH^{\rm eff}_{LL}+\bcalH^{\rm 
eff}_{DD}+\bcalH^{\rm eff}_{RR}$ is the 
uncontacted one-body Hamiltonian of TDDFT while $\bV=\bcalH^{\rm eff}-\bcalE$ 
accounts for the contacting part. 
The fictitious $\bcalG$ will not in general give correct 
one-particle properties. However, by definition, $\bcalG^{<}$ gives 
the correct density
$$
n({\bf r},t)=-2i\bra {\bf r}|\bcalG^{<}(t;t)|{\bf r}\ket 
$$
(where the factor of 2 comes from spin).  
Also total currents are correctly given by TDDFT. If for instance 
$J_{\a}$ is the total current from a particular region $\a$ we have  
\begin{equation}
J_{\a}(t)=-e\int_{\a}d {\bf r}\;\frac{d}{dt}n({\bf r},t) ,
\label{cudft}
\end{equation}
where the space integral extends over the region $\a$ ($e$ is the 
electron charge). 
The TDDFT gives the exact density, and from eq. (\ref{cudft})
we see that it also gives the exact longitudinal current density and total
current from subregions.

Since $\bV_{LR}=\bV_{RL}=0$, from eq. (\ref{eomg}) and 
eq. (\ref{cudft}) the current from the $\a=L,R$ electrode to the 
central region can be written as
$$
J_{\a}(t)=2e\;\Re\left[{\rm tr}
\left\{\bcalG^{<}(t;t)\bV_{\a D}
\right\}\right],\quad\a=L,R.
$$
Without loss of generality we may assume that the partition-free system is 
in equilibrium when $t\leq 0$, \textit{i.e.}, $\bcalE(t\leq 0)\equiv 
\bcalE^{0}$. For the noninteracting system of TDDFT
everything is known once we know how to propagate 
the one-electron orbitals in time and how they are populated
before the system is perturbed. The time evolution is fully
described by the retarded or advanced Green functions $\bcalG^{\rm R,A}$,
and the initial population at zero time, \textit{ i.e.}, 
by $\bcalG^<(0,0)=if(\bcalE^{0}+\bV)$, where $f$ is the Fermi 
distribution function [since $\bcalE^{0}+\bV$ is a matrix, so is 
$f(\bcalE^{0}+\bV)$]. Then, for any $t,t'>0$ we 
have\cite{blandin,cini,sa}
\begin{equation}
\bcalG^{<}(t,t')=
\bcalG^{\rm R}(t,0) \bcalG^<(0,0)\bcalG^{\rm A}(0,t')=i\,
\bcalG^{\rm R}(t,0) f(\bcalE^{0}+\bV)\bcalG^{\rm A}(0,t'),
\label{g<}
\end{equation}
and the total time-dependent current can be rewritten as  
\begin{equation}
J_{\a}(t)=-2e\;\Im\left[{\rm tr}\left\{
\bcalG^{\rm R}(t,0) f(\bcalE^{0}+\bV)\bcalG^{\rm A}(0,t)
\bV_{\a D}
\right\}\right].
\label{current2}
\end{equation}

The above equation is an \textit{exact} result.  
For noninteracting electrons, eq. (\ref{current2}) agrees with the formula 
obtained by Cini\cite{cini} in the 
partition-free scheme. Indeed, the derivation by Cini does not depend 
on the details of the noninteracting system and therefore it is also 
correct for the Kohn-Sham system, which however has the extra 
merit of reproducing the exact density.
The advantage
of this approach is that the interaction in the leads and in the
conductor are treated on the same footing via self-consistent
calculations on the current-carrying system. It also allows for
detailed studies of how the contacts influence the conductance
properties. We note in passing that eq. (\ref{current2}) is also 
gauge invariant since it does not change under an overall 
time-dependent shift of the external potential which is constant in space. 
Neither, is it modified by a simultaneous shift of the classical 
electrostatic potential and $\m$ for $t<0$. 

\section{Asymptotic current}
Once the effective potential of TDDFT is known, the problem is reduced 
to a fictitious one-particle problem. For any such one-particle 
problem we now show that the partitioned scheme by Caroli \textit{et 
al.} and the partition-free scheme are equivalent in 
the long-time limit. To this end, we consider the Dyson equation that 
relates a noninteracting system contacted all the time (Green 
function $\bcalG$) to the corresponding uncontacted one (Green 
function $\bg$). Both systems are exposed to the same, possibly 
time-dependent, potential for $t>0$.
As is well known, the noninteracting Green function $\bcalG$ projected 
in a subregion $\a=L,R$ or $D$ can be described in terms of self 
energies which account for the hopping in and out of the subregion in 
question. Considering the device Green function, the self energy can 
be written $\S^{\rm X}=\sum_{\a=L,R}\S^{\rm X}_{\a}$, with ${\rm 
X=R,A,<}$ or $>$. Here, $\S^{\rm X}_{\a}$ is given by 
$$
\S^{\rm X}_{\a}(t;t')=\bV_{D\a}\,\bg^{\rm X}(t;t')
\bV_{\a D},
$$
where  $\bg$ satisfies eq. (\ref{eomg}) with $\bV=0$. 
The $\bg$ can be expressed in terms of 
the one-body evolution operator $\bcalS(t)$ which fulfils 
$i\dot{\bcalS}(t)=\bcalE(t)\bcalS(t)$ and $\bcalS(0)=1$. The retarded 
and advanced components are $
\bg^{\rm R,A}(t;t')=\mp i\Theta(\pm t\mp t')
\bcalS(t)\bcalS^{\dag}(t')
$. Since also the uncontacted system is initially in equilibrium we have 
$\bg^{<}(t;t')=i\bg^{\rm R}(t;0)f(\bcalE^{0})\bg^{\rm A}(0;t')$ 
[\textit{cf}. eq. (\ref{g<})].

The two schemes are equivalent if  
\begin{equation}
\lim_{t\ra\inf}\S_{\a}^{\rm R}(t;t')=
\lim_{t\ra\inf}\S_{\a}^{\rm A}(t';t)=0
\label{limco}
\end{equation}
for any non-singular $\bV$. Indeed, from the 
equation of motion (\ref{eomg}) and eq. (\ref{limco}) 
one can verify that the projected Green functions satisfy 
\begin{equation}
\lim_{t\ra\inf}\bcalG^{\rm R}_{DD}(t;t') =
\lim_{t\ra\inf}\bcalG^{\rm A}_{D\a}(t';t)\bV_{\a D}=0.
\label{asygra}
\end{equation}
Making use of the above relations one finds\cite{sa} that 
the asymptotic ($t\ra\inf$) time-dependent current of 
eq. (\ref{current2}) becomes 
\begin{equation}
J_{\a}(t)=-2e\;\Im\left[
{\rm tr}
\left\{
\bcalG^{\rm R}(t,0) f(\bcalE^{0})\bcalG^{\rm A}(0,t)
\bV_{\a D}
\right\}
\right].
\label{currcar}
\end{equation}
Thus, the long-time limit washes out the initial effect induced 
by the conducting term $\bV$. Moreover, due to eq. (\ref{asygra}), 
the asymptotic current is independent of the initial equilibrium 
distribution of the central device. 
Expressing $\bcalG^{\rm R}_{D\b}$ and $\bcalG^{\rm A}_{\b\a}$ in 
terms of $\bcalG^{\rm R}_{DD}$ and $\bcalG^{\rm A}_{DD}$ respectively,  
eq. (\ref{currcar}) can be rewritten in a more familiar form
\begin{equation}
J_{\a}(t)=2e\int d{\bar t}\;
\Re\left[
{\rm tr}\left\{
\bcalG^{\rm R}_{DD}(t,{\bar t})\S_{\a}^{<}({\bar t},t)+
\bcalG^{<}_{DD}(t,{\bar t})\S_{\a}^{\rm A}({\bar t},t)
\right\}
\right],
\label{current3} 
\end{equation}
where the asymptotic relation ($t,t'\ra\inf$)
$
\bcalG^{<}_{DD}(t;t')=
\int d{\bar t}\;d{\bar t}'\;\bcalG^{\rm R}_{DD}(t,{\bar t})
\S^{<}({\bar t},{\bar t}')\bcalG^{\rm A}_{DD}({\bar t}',t')
$
has been used. Eq. (\ref{current3}) is valid for interacting
devices connected to interacting electrodes. It provides
a useful framework for studying the transport in interacting
systems from first principles. Our results can
be applied both to the case of a constant (d.c.) bias as
well as to the case of a \textit{time-dependent} 
(\textit{e.g.}, a.c.) one.

To summarise, if the retarded/advanced self energy satisfies 
eq. (\ref{limco}), the exact formula in eq. (\ref{current2}) 
reduces to eq. (\ref{current3}) in the long-time limit. For 
noninteracting electrons the Green function $\bcalG$ of TDDFT 
coincides with the Green function of the real system and 
eq. (\ref{current3}) agrees with the formula by Wingreen \textit{et al.} 
\cite{wingreen,jauho}

\section{Steady state}
Next, we wish to investigate the steady state, \textit{i.e.}, 
$\lim_{t\ra\inf}\bcalE(t)=\bcalE^{\inf}=$ const. In this case it must 
exist a unitary operator $\bar{\bcalS}$ such that
$
\lim_{t\ra\inf}\bcalS(t)={\rm e}^{-i\bcalE^{\inf}t}\;\bar{\bcalS}.
$
Then, in terms of diagonalising 
one-body states $|\f^{\inf}_{m\a}\ket$ of $\bcalE^{\inf}_{\a\a}$ with 
eigenvalues $e^{\inf}_{m\a}$ we have 
$$
\S^{<}_{\a}(t;t')=i\sum_{m,m'}
{\rm e}^{-i[e^{\inf}_{m\a}t-e^{\inf}_{m'\a}t']}
\bV_{D\a}|\f^{\inf}_{m\a}\ket \bra\f^{\inf}_{m\a}|
f(\bar{\bcalE}^{0})|\f^{\inf}_{m'\a}\ket\bra\f^{\inf}_{m'\a}|
\bV_{\a D},
$$
where $\bar{\bcalE}^{0}=\bar{\bcalS}\;\bcalE^{0}\bar{\bcalS}^{\dag}$.
For $t,t'\ra\inf$, the left and right contraction with a nonsingular $\bV$ 
causes a perfect destructive interference for states with 
$|e^{\inf}_{m\a}-e^{\inf}_{m'\a}| \gtrsim 1/(t+t')$ and hence the 
restoration  
of translational invariance in time
\begin{equation}
\S^{<}_{\a}(t;t')=i\sum_{m}f_{m\a}\G_{m\a}{\rm e}^{-ie^{\inf}_{m\a}(t-t')},
\label{asyse}
\end{equation}
where $f_{m\a}=\bra\f^{\inf}_{m\a}|f(\bar{\bcalE}^{0})|\f^{\inf}_{m\a}\ket$ 
while 
$
\G_{m\a}=\bV_{D\a}|\f^{\inf}_{m\a}\ket 
\bra\f^{\inf}_{m\a}|
\bV_{\a D}$ \cite{endnote}.
The above \textit{dephasing mechanism} is the key ingredient in the 
developing of a steady state. Substituting eq. (\ref{asyse}) into 
eq. (\ref{current3}) we get the steady state current 
\begin{equation}
J_{\a}=-2e\sum_{m\b}f_{m\b}\left[\d_{\a\b}{\rm tr}\left\{\G_{m\a}
\Im[\bcalG^{\rm R}_{DD}(e^{\inf}_{m\a})]
\right\}+{\rm tr}\left\{
\bcalG^{\rm R}_{DD}(e^{\inf}_{m\b})\G_{m\b}
\bcalG^{\rm A}_{DD}(e^{\inf}_{m\b})
\Im[\S^{\rm A}_{\a}(e^{\inf}_{m\b})]
\right\}\right]
\label{current4}
\end{equation}
with
$
\bcalG^{\rm R,A}_{DD}(\ve)=[\ve-\bcalE^{\inf}_{DD}-\S^{\rm 
R,A}(\ve)]^{-1}.
$
Using the equalities $\Im[\bcalG^{\rm R}_{DD}]=
[\bcalG^{\rm R}_{DD}-\bcalG^{\rm A}_{DD}]/2i$ and 
$[\bcalG^{\rm R}_{DD}-\bcalG^{\rm A}_{DD}]=
[\bcalG^{>}_{DD}-\bcalG^{<}_{DD}]$ together with
$$
[\bcalG^{>}_{DD}(\ve)-\bcalG^{<}_{DD}(\ve)]=
-2\p i\sum_{m\a}\d(\ve-e^{\inf}_{m\a})\bcalG^{\rm R}_{DD}(e^{\inf}_{m\a})
\G_{m\a}\bcalG^{\rm A}_{DD}(e^{\inf}_{m\a})
$$
and
$
\Im[\S^{\rm A}_{\a}(\ve)]=\p \sum_{m}\d(\ve-e^{\inf}_{m\a})
\G_{m\a},
$
the steady-state current in eq. (\ref{current4}) can be rewritten in 
a Landauer-like form 
\begin{equation}
J_{R}=-e\sum_{m}[f_{mL}{\cal T}_{mL}-f_{mR}{\cal T}_{mR}]=-J_{L}.
\label{current5}
\end{equation}
In the above formula ${\cal T}_{mR}=\sum_{n}{\cal T}_{mR}^{nL}$ and 
${\cal T}_{mL}=\sum_{n}{\cal T}_{mL}^{nR}$ are the TDDFT transmission coefficients
expressed in terms of the quantities 
$$
{\cal T}_{m\a}^{n\b}=
2\p\d(e^{\inf}_{m\a}-e^{\inf}_{n\b}){\rm tr}\left\{
\bcalG^{\rm R}_{DD}(e^{\inf}_{m\a})\G_{m\a}
\bcalG^{\rm A}_{DD}(e^{\inf}_{n\b})\G_{n\b}
\right\}={\cal T}_{n\b}^{m\a}.
$$
Despite the formal analogy with the Landauer formula, 
eq. (\ref{current5}) contains an important conceptual difference since
$f_{m\a}$ is not simply 
given by the Fermi distribution function. For example, if the 
induced change in effective potential varies widely in space 
deep inside the electrodes, the band structure $\bar{\bcalE}^{0}_{\a\a}$
may be completely different from that of 
$\bcalE^{\inf}_{\a\a}$. However, if 
we asymptotically have equilibrium 
far away from the central region, as we would expect for leads with a 
macroscopic cross section, the change in effective potential must be 
uniform. To leading order in $1/N$ we then have  
\begin{equation}
\bcalE_{\a\a}(t)=\bcalE^{0}_{\a\a}+\D U^{\rm eff}_{\a}(t),
\label{asytddft}
\end{equation}
and $\bcalE^{\inf}_{\a\a}=\bcalE^{0}_{\a\a}+\D U^{\rm eff}_{\a,\inf}$. 
Hence, 
except for corrections which are of lower order 
with respect to the system size, 
$\bar{\bcalE}^{0}_{\a\a}=\bcalE^{0}_{\a\a}$ and 
$$
f_{m\a}=f(e^{\inf}_{m\a}-\D U^{\rm eff}_{\a,\inf}).
$$  

We emphasise that the steady-state current in 
eq. (\ref{current5}) comes out from a pure dephasing mechanism in the 
fictitious noninteracting problem. 
The damping effects from scatterings is described by 
$A_{\rm xc}$ and $v_{\rm xc}$. Furthermore, if eq. (\ref{asytddft}) 
holds, the current depends only on the asymptotic value of the 
effective potential, $U^{\rm eff}({\bf r},t\ra\inf)$. 
However, $U^{\rm eff}({\bf r},t\ra\inf)$ may depend on the history of the 
external bias $U$. Thus, the steady state 
current of fully interacting systems may be history dependent.
In the case of Time Dependent Local Density 
Approximation, the exchange-correlation potential $v_{\rm xc}$ 
depends only locally on the instantaneous density and has no memory at all. 
If the density tends to a constant, so does the effective potential 
$U^{\rm eff}$, which again 
implies that the density tends to a constant. Owing to the non-linearity 
of the problem there might still be more than one steady-state 
solution or none at all.

The Landauer formula in eq. (\ref{current5}) corresponds closely 
to the result obtained by
Lang and coworkers \cite{Lang1,Lang2}.
In their approach, the continuum
is split into left- and right-going parts which are populated
according to two different chemical potentials. The density
is then calculated self-consistently. In our language this 
corresponds to writing the uncontacted $\bg^{<}(\ve)$ as 
$$
\bg^{<}(\ve)=2\p i \sum_{m\a}f_{\a}(e^{\inf}_{m\a})
\d(\ve-e_{m\a}^{\inf})
|\f^{\inf}_{m\a}\ket\bra \f^{\inf}_{m\a}| 
$$
in terms of Fermi functions $f_{\a}$ with chemical potential 
$\m_{\a}=\m+\D U^{\rm eff}_{\a,\inf}$.
The chemical potentials for the two leads differ, and the final result
is independent of the chosen chemical potential for the device due to 
eq. (\ref{limco}).
When we apply $1 + \bcalG^{\rm R}\bV=
\bcalG^{\rm R}[\bg^{\rm R}]^{-1}$ to an unperturbed state 
$|\phi_{m\alpha}^{\inf}\ket$, it is transformed to an interacting, 
\textit{i.e.}, contacted eigenstate $|\psi_{m\alpha}^{\inf}\ket$. 
Above the conductance threshold, 
states originating from the left lead become right-going scattering
states, and states from the right lead become left-going scattering
states. In addition, fully reflected waves and discrete states may arise
which contribute to the density but not to the current.
Thus,
$$
\bcalG^{<}(\ve)= [ 1 + \bcalG^{\rm R}(\ve)\bV]\; \bg^<(\ve) 
[  1 + \bV\bcalG^{\rm A}(\ve) ]=
2\p i \sum_{m\a}f_{\a}(e^{\inf}_{m\a})
\d(\ve-e_{m\a}^{\inf})
|\q^{\inf}_{m\a}\ket\bra \q^{\inf}_{m\a}| .
$$

Lang \textit{et al.} further
approximate exchange and correlation by the LDA and the leads
by homogeneous jellia, but apart from these approximations it is
clear that their method implements TDDFT, as described
above, in the steady state. 
It is also clear that the correctness of the 
Lang's approach relies on eq. (\ref{limco}) and eq. (\ref{asytddft}), as 
we have seen before. The equivalence between the scattering state
approach by Lang \textit{et al.} and the partitioned non-equilibrium
approach used by Taylor \textit{et al.} \cite{Taylor1,Taylor2}
has also been shown by Brandbyge \textit{et al.} \cite{brand}
\section{Conclusions}
In conclusion, we have proposed a partition-free scheme based on TDDFT 
in order to treat the time-dependent current response of fully 
interacting systems. Our formally exact theory is 
more akin to the way the experiments are carried out and 
allows us to calculate 
the physical (\textit{i.e.}, measurable) 
dynamical current responses. Among the advantages 
we stress the possibility to include the interactions between 
the leads and between the leads and the device in a very natural way. 
We have shown that the steady state develops due to a 
\textit{dephasing mechanism} without any reference to many-body damping and 
interactions.  The damping 
mechanism (due to the electron-electron scatterings) of the real 
problem is described by $v_{\rm xc}$. The steady-state current depends 
on the history only through the asymptotic shape of the effective TDDFT 
potential $U^{\rm eff}$ provided the bias-induced change  
$\D U_{\a}^{\rm eff}$ is uniform deep inside the electrodes. (This is 
the anticipated behaviour for macroscopic electrodes.)
In such systems, we have shown that the nonlinear steady-state current 
can be expressed in a Landauer-like formula in terms of fictitious 
transmission coefficients and one-particle energy eigenvalues. Our 
scheme is equally applicable to time-dependent responses. Clearly, its usefulness
depends on the quality of the approximate TDDFT functionals 
being used. We are presently investigating how different 
approximations influence the current response in model systems.

\acknowledgments
We would like to acknowledge useful discussions with
U. von Barth, 
P. Bokes, 
M. Cini, R. Godby, A.-P. Jahuo, B. I. Lundqvist, P. Hyldgaard,
and B. Tobiyaszewska.
This work was supported by the RTN program of the European Union
NANOPHASE (contract HPRN-CT-2000-00167).


\begin{thebibliography}{0}
\bibitem{endnote}
In principle, there may be degeneracies which require a diagonalisation
to be performed for states on the energy shell.
\end{thebibliography}
\end{document}